\documentclass[showpacs,twocolumn]{revtex4}
\usepackage{amsmath,amsfonts,amssymb}
\usepackage[dvips]{graphicx}
\def\lm {\lambda}
\def\rs {\rho_s}

\begin{document} 
\title{Equilibrium adsorption on a random site surface}
\author{J. Talbot$^1$, G. Tarjus$^2$ and P. Viot$^2$}
\affiliation{$^1$Department   of  Chemistry  and  Biochemistry,
 Duquesne University, Pittsburgh, PA 15282-1530\\ $^2$Laboratoire de
 Physique Th{\'e}orique de la Mati\`ere Condens\'ee, Universit{\'e} Pierre et
 Marie Curie, 4, place Jussieu,75252 Paris Cedex, 05 France}

\begin{abstract}
We examine the reversible adsorption  of spherical solutes on a random
site surface in which the  adsorption sites are uniformly and randomly
distributed on  a substrate.  Each site can  be occupied by one solute
provided  that the nearest  occupied  site is  at  least one  diameter
away.  The model  is characterized by   the site density  and the bulk
phase activity  of the  adsorbate.   We develop a  general statistical
mechanical  description of the model and   we obtain exact expressions
for  the adsorption  isotherms in limiting  cases  of  large and small
activity and   site  density, particularly   for  the  one dimensional
version of  the model.  We   also propose approximate  isotherms  that
interpolate   between the exact results.  These   theories are in good
agreement with numerical simulations of the model in two dimensions.
\end{abstract}

\maketitle

\section{Introduction}

Heterogeneity often   plays  an important role   in various adsorption
processes  and its   presence may  profoundly   modify  the adsorption
isotherms  and    other   thermodynamic properties   compared   to the
homogeneous          situation.        Consequently,          numerous
articles\cite{B04,ASWM02},   reviews     \cite{D01,Adamczyk2005}   and
monographs  \cite{JM88,RE92}   have   been   published that   describe
experimental, theoretical  and   numerical studies in  the   area. The
disorder may originate from the  energetic or structural heterogeneity
of the substrate or from the adsorbate species. Heterogeneity may also
result  if the adsorbed molecules are   large enough so that multisite
adsorption becomes possible \cite{JWA96}.

In a recent article \cite{TTV07},  we investigated the properties of a
lattice  model of adsorption  on a  disordered  substrate that can  be
solved  exactly (See  also Refs.\cite{Oshanin2003,OBB03}).   We showed
that there exists  an exact mapping  to the system without disorder in
the  limits  of small and infinite   activities  and we exploited this
result  to   obtain an approximate,   but  accurate description of the
disordered system.  

For continuous systems, structural  disorder may be represented by the
random site model (RSM)  in which adsorption  sites are uniformly  and
randomly distributed  on  a plane  \cite{JWTT93,OT07}. The  molecules,
represented by hard spheres, can bind to these immobile sites.  Steric
exclusion  is expressed  by the   fact that  a  site  is available for
adsorption only if the nearest occupied  site is at least one particle
diameter  away.  In addition,  adsorption  energy is assumed equal for
each  adsorbed molecule.   Therefore,  the disorder of  this model  is
characterized by the dimensionless site  density, $\rho_s$, only.  The
degree   of complexity increases  drastically  because steric effects,
which usually dominate    the adsorption on continuous   surfaces, are
modified by the local disordered  structure of adsorption sites.  This
model may be appropriate for  the reversible adsorption of proteins on
disordered substrates\cite{OGMRW03,LJ05}.

Oleyar   and Talbot \cite{OT07} examined the
reversible  adsorption of hard spheres on  the 2D RSM  and proposed an
approximate theory  for the  adsorption isotherms  based  on a cluster
expansion  of    the grand   canonical  partition   function. Although
successful  at  low site    densities,   the quality  of  the   theory
deteriorates rapidly with increasing site density and fails completely
above a certain density.

In sections II and  III we develop  a general statistical  mechanical
description of the RSM model and we  confirm the intuitive result that
in  the limit of  large site density the system   maps to hard spheres
adsorbing  on a continuous surface. Then,  for finite site density, we
prove that  for the  one-dimensional system in  the limit  of infinite
activity, there is a mapping to a hard rod system at a pressure $\beta
P=\rs$. We  also present an argument supporting  the validity  of this
result in  higher  dimensions.    Note that  when the    activity   is
``strictly''   infinite, desorption is  no    longer possible and  the
adsorption process is irreversible.  In  this case  the model has   an
exact mapping   to  the Random   Sequential Adsorption (RSA)   of hard
particles on a continuous surface
\cite{JWTT93}.

In sections  IV and V we propose  approximate theoretical  schemes for
the adsorption isotherms  in one and  two  dimensions that interpolate
between the limits  of small  and  large activities for a  given  site
density $\rho_s$. Comparison  with simulation results shows that these
approaches are a considerable improvement over the cluster expansion.

For completeness we note that in addition to its application to
adsorption, the model is also interesting because of its relationship
to the vertex cover problem\cite{WH00,WH01}.  A vertex cover of an
undirected graph is a subset of the vertices of the graph which
contains at least one of the two endpoints of each edge. In the vertex
cover problem one seeks the {\it minimal vertex cover} or the vertex
cover of minimum size of the graph.  This is an NP-complete problem
meaning that it is unlikely that there is an efficient algorithm to
solve it.  The connection to the adsorption model is made by
associating a vertex with each adsorption site.  An edge is present
between any two vertices (or sites) if they are closer than the
adsorbing particle diameter.  The minimal vertex cover corresponds to
densest particle packings.  Weight and Hartmann\cite{WH00,WH01}
obtained an analytical solution for the densest packing of hard
spheres on random graphs, but the existence of the geometry in
adsorption processes implies that the machinery developed to describe
adsorption on random graphs cannot be used in RSM models.

\section{Statistical Mechanics of the Random Site Model }

The adsorption surface is generated by placing $n_s$ points,
representing adsorption sites, randomly and uniformly on a substrate,
either a line in 1D or a plane in 2D (the boundary conditions are
irrelevant in the large $n_s$ limit).  Spheres of diameter $\sigma$ may
bind, centered, on an available adsorption site.  A site is available
if the nearest occupied site is at least a distance $\sigma$ away.  Two
points are connected, and therefore cannot be simultaneously occupied,
if they are closer than $\sigma$.

The positions of  the $n_s$ sites are denoted  by ${\bf R}_i$ where $i$ is a
index running  from $1$ to $n_s$. The   sites are quenched  during the
adsorption-desorption process.  The adsorbed phase in equilibrium with
a bulk phase  containing  adsorbate  at   activity $\lambda$  can   be
formally described with the grand canonical partition function:

\begin{equation}\label{eq:pf}
\Xi(\lambda,\{{\bf R}_i\})= 1+\sum_{n=1}^{\infty}\frac{\lambda^n}{n!}
\int\cdots\int d{\bf r}^n\prod_{i>j}(1+f_{ij})\prod_{i=1}^n\eta({\bf r}_i)
\end{equation}
where  the microscopic density of sites $\eta({\bf r})$  is given by
\begin{equation}\label{eq:7}
\eta({\bf r})=\sum_{i=1}^{n_s}\delta({\bf r}-{\bf R}_i),
\end{equation}
$\lambda=\exp(\beta\mu)$  is   the activity, ${\bf   r}_i$ denotes the
position of  sphere $i$ and $f_{ij}$  is the Mayer f-function which is
equal to $-1$ if spheres $i$ and $j$ are closer  than $\sigma$ and $0$
otherwise.  Eq.  (\ref{eq:pf}) applies to  a particular realization of
the quenched  variables $\{R_i\}$. In  order to average over disorder,
it is necessary to take the average not of the partition function, but
of the logarithm of  the partition function  \cite{RST94}.  By  using
standard rules of diagram theory\cite{HM76}, one obtains that
\begin{align}\label{eq:11}
\overline{\ln(\Xi)}=\sum_{n=1}^{\infty}\frac{\lambda^n}{n!}
\int d{\bf r}^n U_n({\bf r}_1,{\bf r}_2,...,{\bf r}_n)\prod_{i=1}^n\overline{\eta({\bf r}_i)}
\end{align}
where  the   bar  means    that      the  average   is   taken    over
disorder,  $U_n({\bf   r}_1,{\bf r}_2,...,{\bf   r}_n)$
denotes the Ursell  function associated with the Mayer f-functions of  hard
particles, $\prod_{i>j}(1+f_{ij})$\cite{G76}. 

 Let us denote the probability of finding  adsorption 
sites at positions ${\bf R_i},i=1,...,n_s$, as
$P({\bf R}_1,{\bf R}_2,...{\bf R}_s)$. We will assume that the positions are uncorrelated so that
\begin{equation}\label{eq:15}
P({\bf R}_1,{\bf R}_2,...{\bf R}_s)=\prod_{i=1}^{n_s}P({\bf R}_i)
\end{equation}
and will consider a Poissonian distribution of points,
$P({\bf R})=1/A$.
The average of the site density $\overline{\eta({\bf r})}$ is given by
\begin{align}\label{eq:13}
\overline{\eta({\bf r})}&=\int...\int\prod_{i=1}^{n_s}d{\bf R}_i P({\bf R}_1,{\bf R}_2,...{\bf R}_s)\eta({\bf r})\nonumber\\
&=\rho_s
\end{align}
where $\rho_s=n_s/A$ is the site density of the particles in a system
of area $A$ (length in one dimension). 

We show in Appendix A that the average of the logarithm of the partition function over disorder can be written as
\begin{equation}
\overline{\ln(\Xi)}= \ln(\Xi^*(z=\lambda\rho_s)) +O(1/\rho_s)
\end{equation}
where 
\begin{align}\label{eq:12}
\ln(\Xi^*(z))&=\sum_{n=1}^{\infty}\frac{z^n}{n!}
\int d{\bf r}^n U_n({\bf r}_1,{\bf r}_2,...,{\bf r}_n)
\end{align}
i.e., $\ln(\Xi^*(z))$, is  the partition  function  of hard spheres on  a
continuous surface  at  an  activity $z=\lambda\rho_s$.  This   result
shows that when  the site density  $\rho_s$ goes to infinity and  that
the activity  $\lambda$  goes  to  $0$, with the   constraint that the
product $\lambda\rho_s$ remains finite,  the Random Site Model maps to
a system of hard particles in continuous space.

The  number density of adsorbed molecules 
can be computed directly from the partition function:
\begin{equation}
\rho(\lambda,\rho_s)=\frac{z}{A}\left(\frac{\partial\overline{\ln \Xi}}{\partial z}\right)_{n_s}
\end{equation}
with again $z=\lambda\rho_s$.

By using Eq.(\ref{eq:21}) given in the Appendix, one obtains to the second-order in $1/\rho_s$
\begin{equation}
\rho(\lambda,\rho_s)=\rho^*(z)-\frac{\rho^*(z)}{\rho_s+\rho^*(z)}\frac{d\rho^*(z)}{dz}+O(1/\rho_s^2)
\end{equation}
where   $\rho^*(z)$ is the number  density   of the hard sphere model in continuous space.
Since $\rho^*(z)$ is an increasing function of the activity $z$, it is
an  upper   bound for   $\rho(\lambda,\rho_s)$.

The expansion of $\overline{\ln \Xi}$ and $\rho(\lambda,\rho_s)$ in powers of the activity when $\lambda\rightarrow 0$ can also be generated from the above equations in straightforward way.
\section{The limit of large activity}
\subsection{A mapping with the (homogeneous) equilibrium hard sphere model}
The limit of large activity, $\lambda\rightarrow\infty$, is expected
to lead to the maximum density of adsorbed spheres for a given density
$\rho_s$ of adsorption sites. Indeed, this limit combines the presence
of a relaxation mechanism, which is induced by the infinitesimally small but
non-zero desorption process that allows a sampling of hard-sphere
configurations, with the guarantee that no sites left open for
adsorption will stay empty. The one-dimensional model is then amenable
to an exact solution in the limit
$\lambda\rightarrow\infty$. Interestingly, it maps onto an equilibrium
system of hard rods in continuum (1D) space at the same density (which
is of course a function of $\rho_s$)

The adsorbed density $\rho(\rho_s,\lambda\rightarrow\infty)=\rho_{\rm
max}(\rho_s)$ in the 1D case can be obtained exactly with the
following simple probabilistic argument. Assume  that  a given  site is
occupied.   Then  all sites that  lie within  a distance $\sigma$ from
this site must be unoccupied.  In  the the optimally packed system the
next site beyond this must be  occupied.  The average distance from an
arbitrary  point to the first  site  is $1/\rho_s$  giving the average
distance between two occupied  sites as $\sigma+1/\rho_s$. The average
coverage in the maximally occupied system is thus
\begin{equation}\label{eq:rmax}
\rho_{\rm max}(\rho_s)=\frac{\rho_s}{1+\rho_s\sigma}.
\end{equation}
In the following we set $\sigma=1$. When the site density is low, $\rho_s<<1$, one recovers the
independent site approximation (Langmuir model), $\rho_{\rm
max}=\rho_s$. As the site density increases Eq. (\ref{eq:rmax}) shows
that there is a continuous increase of $\rho_{\rm max}$ and that a
closed packed configuration is obtained as $\rho_s$ approaches
infinity

The above  argument can be generalized
to describe the correlation functions, or more conveniently in this 1D
system, the gap distribution functions. Specifically let 
$F(x)$ denote the
probability density associated with finding a gap of size between $x$
and $x+dx$. In an optimally packed configuration, the probability to
find a gap of length $x$ is related to the probability to find the
first site at a given distance $x$ (say to the right) of an arbitrary
point. For a Poissonian distribution of sites this simply leads to
\begin{equation}\label{eq:fmax}
F_{\rm max}(x;\rho_s)=\rs e^{-\rs x}. 
\end{equation}
The reasoning is easily extended  to multi-gap distribution functions,
$F(x_1,x_2),F(x_1,x_2,x_3),...$,      where two   successive  gaps are
neighbors   in the   sense that    they  are   separated  by a   single
particle. One  then  shows that all  these higher-order  gap functions
factorize in products of one-gap distribution functions, e.g. $ F_{\rm
max}(x_1,x_2) = F_{\rm max}(x_1)F_{\rm max}(x_2)$. The outcome of this
probabilistic argument is that the 1D random  site model is equivalent
to an  equilibrium system of hard  rods on a  (continuous) line at the
pressure
\begin{equation}\label{eq:bp}
\beta P=\rho_s.
\end{equation}
Indeed,   from the  known equation  of  state  of  the  hard rod fluid
\cite{T36}  one   has    $\beta  P=\rho/(1-\rho)$,  i.e.   $\rho=\beta
P/(1+\beta   P)$, which corresponds    to  Eq.  (\ref{eq:rmax})  after
insertion of   Eq. (\ref{eq:bp}), and $F(x)  =  \beta Pe^{-\beta Px}$,
which reduces again to Eq. (\ref{eq:fmax}). The multi-gap distribution
functions are also given  by products of 1-gap distribution functions,
which completes the proof of equivalence.

Extension of the above arguments to higher dimensions is far from
straightforward. First, in $d=2$ and higher, one may encounter at high
adsorbed density phase transitions to ordered, crystalline-like,
phases. Second, the simplicity of the reasoning in terms of gaps
characterized only by their length is lost when one leaves the
one-dimensional case. For these reasons, we have not been able to
develop a rigorous demonstration of a mapping between the 2D RSM and
the (homogeneous) hard-disk system at equilibrium at the same density
when the activity $\lambda$ goes to infinity.

To nonetheless  make   some progress, let   us consider   the  nearest
neighbor radial distribution   function $H(r)$ introduced  by Torquato
\cite{T95}.  $H(r)$   (and also used   in  the context of irreversible
adsorption   models\cite{RTT96,VTT98}   is    the  probability density
associated   with finding a nearest neighbor   particle center at some
radial distance $r$ from the reference particle center.  This somewhat
generalizes  the    1-gap  distribution  function   to  any    spatial
dimension. For   a  Poissonian  distribution of particle    centers of
density $\rs$ in $d=2$, one finds that
\begin{equation}\label{eq:hpoisson}
H(r) = 2\pi r\rs e^{-\pi r^2\rs}
\end{equation}
No exact formula exists for $H(r)$ in a hard disk fluid. Using the results
of reference \cite{T95} one can,
however, demonstrate that in the large $r$ limit at an equilibrium
pressure $P$ it is given by
\begin{equation}\label{eq:hlim}
H(r)\sim 2\pi r\beta P  e^{-\pi r^2\beta P}
\end{equation}
The reasoning now  goes   as follows.  Consider a  maximally  occupied
configuration  (associated with  the limit $\lambda\rightarrow\infty$)
in the  2D RSM with  an adsorption site  density  $\rs$ and consider a
given adsorbed particle. All  sites within a radial distance  $\sigma$
of the central occupied site  must be unoccupied. $H(r)$ is associated
with situations such that the nearest adsorbed particle center is at a
radial  distance $r$ of the  reference one. Therefore, there should be
no adsorption sites in the   region delimited by   the two circles  of
radius $\sigma$  (inside) and $r$  (outside). Actually, this statement
is not quite right: in 2 dimensions, there may  be an exclusion effect
due to other particle centers at a distance $\gtrsim r$ of the central
site.  To be more  rigorous, the outside  circle delimiting the region
with  no  adsorption    sites   should  be  (at    least)   of  radius
$r-\sigma$.    When   $r\rightarrow\infty$,  one    can   neglect  the
contribution  of  width  $\sigma$ due  either  to  the  disk of radius
$\sigma$  centered  on the reference  site  or to  the shell  of width
$\sigma$     located between     $r-\sigma$     and    $r$.       When
$r\rightarrow\infty$,  the probability  that,  given that  a reference
particle is centered at  the origin, a  spherical region of radius $r$
is empty of adsorption sites is  given by $\exp(-\rs\pi r^2)$, so that
asymptotically $H_{\rm  max}(r)$  goes as in  Eq. (\ref{eq:hpoisson}),
i.e.
\begin{equation}
H_{\rm max}(r)\sim 2\pi r\rs  e^{-\pi r^2\rs}
\end{equation}
Comparison with Eq. (\ref{eq:hlim}) tells us that it is the same
asymptotic behavior as that of an equilibrium hard disk system in the
plane with $\beta P=\rs$. This is the same result as found in $d=1$
(see above), except that the density $\rho$ and pressure $P$ are no
longer related by a simple analytical expression.

\begin{figure}[t]
\resizebox{8cm}{!}{\includegraphics{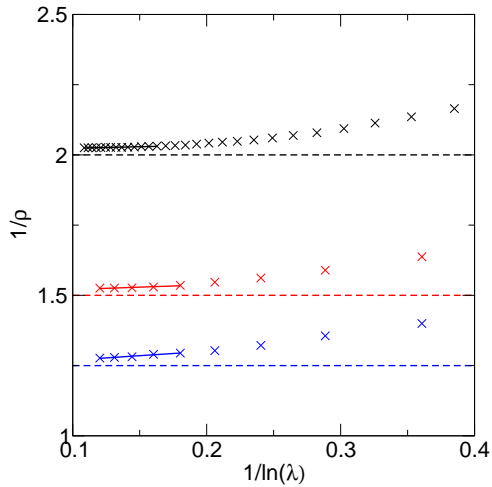}}
\caption{Simulated adsorption Isotherms. $\rho_s=1,2,4$ top to bottom. The dashed lines 
show the predictions of Eq. (\ref{eq:rmax}).}\label{fig:isotherms}
\end{figure}

\subsection{Numerical verification of the suggested mapping}

\begin{figure}[th]

\resizebox{8cm}{!}{\includegraphics{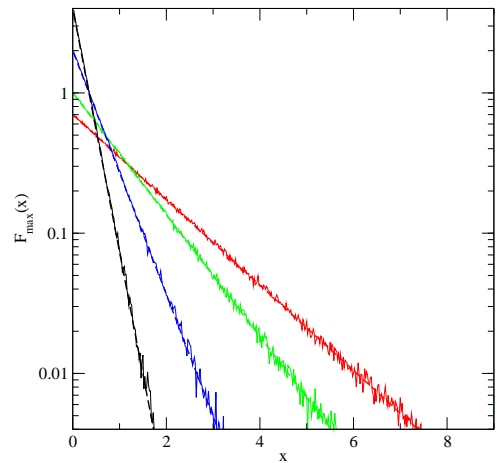}}
\caption{Gap distribution function $F_{\rm max}(x)$ for configurations of hard rods at infinite activity
 on the random site surface with $\rs  = 0.7,1,2.0, 4.0$ from right to
 left,  bottom.  The dashed lines are the predictions of Eq.(\ref{eq:fmax})
 The solid lines correspond to the simulation results.}\label{fig:gapdist1}.
\end{figure}

We  confirmed Eq. (\ref{eq:rmax})  in   two ways.   The first   generates
configurations of maximum density directly.  A number $n_s$ points are
distributed  uniformly and randomly  in the unit  interval.  The first
site is occupied by a rod of length $\sigma=1/ n_s$. Each site to the
right  is checked in  order  until one is  found  that  is at least  a
distance $\sigma$ from the occupied one. This site is occupied and the
process iterated until all sites have been  accounted for. A number of
averages over different configurations of  sites is performed.  We also
verified Eq. (\ref{eq:rmax})  by determining the adsorption isotherms for
different values of  the site  density  $\rho_s$ and taking  the limit
$\lm\to\infty$: See Fig. \ref{fig:isotherms}.

For the hard  rod  fluid at equilibrium the gap distribution function is
given exactly by:

\begin{equation}\label{eq:gdist}
F(\rho,x)=\frac{\rho}{1-\rho}\exp(-\frac{\rho \,x}{1-\rho}).
\end{equation}

The gap distribution function calculated 
for the densest configurations of the
random site model is in agreement with the predictions of Eq.
(\ref{eq:gdist}) using a density computed from Eq. (\ref{eq:rmax}), 
confirming that the   configuration of rods on  the
random site surface has the same structure as the equilibrium hard rod
fluid at  the same density (see Fig.\ref{fig:gapdist1}). This is  consistent with the behavior of a
related  lattice model  for which  we showed  that  there  is an exact
mapping to     the     system   without   disorder       at   infinite
activity\cite{TTV07}.

\begin{figure}[t]
\resizebox{8cm}{!}{\includegraphics{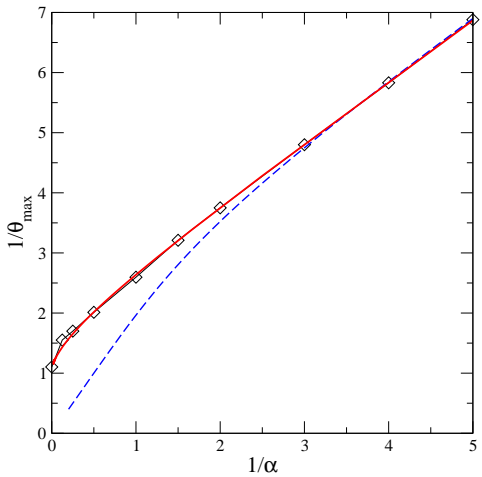}}
\caption{Maximum coverage of the RSM as a function
of the dimensionless  site density. The  symbols show the  simulation
results, the dashed line is  the cluster approximation to second order
\cite{OT07},  and  the   solid    line   shows the   predictions     of
Eqs. (\ref{eq:bp}) and (\ref{eq:wang}). }\label{fig:maxcov2d}.
\end{figure}

In order to use  Eq.(\ref{eq:bp}) to estimate  the maximum coverage of
the RSM in two-dimensions, we combine it with the approximate equation
of state for hard disks on a continuous surface, which was proposed by
Wang\cite{W02}:

\begin{equation}\label{eq:wang}
\frac{\beta P}{\rho}=1+\frac{D}{1-\theta / \theta_0}-(D+2a\theta+4b\theta^2+8c\theta^3+16d\theta^4+32e\theta^5)
\end{equation}
where $\theta=\pi\sigma^2\rho/4$ is the coverage,   $\theta_0=0.907..$
is  the  coverage   of  a   hexagonal   close-packed  configuration,
$D=4.08768, a=1.25366, b=0.46051,  c=0.152797,d=0.04412,  e=0.00929.$.
 For a given  value of the  site density, $\rho_s$,
Eq. (\ref{eq:wang})  is  solved numerically   for $\theta$. It is 
convenient to introduce the dimensionless site density
\begin{equation}\label{eq:14}
\alpha = \frac{\pi}{4}\sigma^2\rho_s
\end{equation}
The  results,  shown in Fig.~\ref{fig:maxcov2d} are in   excellent
agreement with the  simulation    results for  the entire  range    of
$\alpha$. This  should be compared  with  cluster expansion  to second
order       that    gives    good         predictions     only     for
$\alpha\lesssim0.3$\cite{OT07}. Note  that the approximate equation of
state, Eq.(\ref{eq:wang}), does not include the possible presence of a
phase transition.

\begin{figure}[t]
\begin{center}
\resizebox{8cm}{!}{\includegraphics{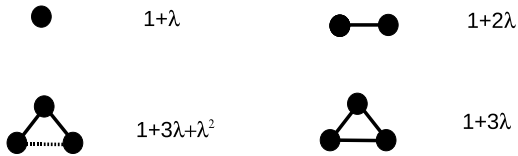}}
\caption{Lowest order clusters of adsorption sites.
A solid line connects  two sites that are closer  than $\sigma$ and  a
dashed   line  indicates  that   two  sites  are  further   apart than
$\sigma$. Triplets of type ``3a'' (left)  and ``3b'' (right) are shown
left to right in  the second row. The expression  to the right  of the
clusters is the corresponding grand  canonical partition function  for
the adsorbed particles.}\label{fig:clusters}
\end{center}
\end{figure}

\section{RSM in one dimension}

\subsection{Low site density expansion}
To   provide  a  description of the    adsorption isotherms  at finite
activity, it  is useful  to consider  an expansion  in the  density of
adsorption  sites.  Assuming  that the adsorption  surface consists of
isolated clusters of sites, the  (averaged) logarithm of the partition
function may be expressed as

\begin{equation}
\overline{\ln \Xi }=N_1\ln \Xi_1 +N_2\ln \Xi_2 +N_{3a}\ln \Xi_{a} +N_{3b}\ln \Xi_{3b}+
...,
\end{equation}
where $N_i=x_iN_s$ is the number of clusters of  type $i$ and $i\equiv
(n,a)$    with $n$ being the  number  of   sites in  the   cluster  and $a$
characterizing  when    necessary  the   subclass  of  clusters   (see
Fig.\ref{fig:clusters}). This  simple  expansion is  possible  because
adsorption on a given cluster does not affect  any of the others.  The
adsorption isotherm is then
\begin{equation}\label{eq:2}
\rho A = \lm N_1\left(\frac{\partial\ln \Xi_1}{\partial\lambda}\right)+\lm N_2\left(\frac{\partial\ln \Xi_2}{\partial\lambda}\right)+...,
\end{equation}

which gives
\begin{align}\label{eq:ce}
\rho &= \lm \rho_s\bigl(x_1\frac{1}{1+\lm}+\frac{2}{1+2\lm}x_2\nonumber\\
+&\frac{3+2\lm}{1+3\lm+\lm^2}x_{3a}+\frac{3}{1+3\lm}x_{3b}+...\bigr).
\end{align}

In one-dimension, exact expressions can be obtained for the
first few clusters following the approach of Quintanilla and Torquato \cite{QT96}:

\begin{align}\label{eq:cex1}
x_1&=\exp(-2\rho_s)\\
\label{eq:cex2}
x_2&=\exp(-2\rho_s)(1-\exp(-\rho_s))\\
\label{eq:cex3a}
x_{3a}&=\exp(-3\rho_s)(\exp(-\rho_s)+\rho_s-1)\\
\label{eq:cex3b}
x_{3b}&=\exp(-2\rho_s)(1-(1+\rho_s)\exp(-\rho_s))
\end{align}
where $x_2$,  for example, is the number  of clusters  with two linked
sites (per adsorption site). At the triplet level we need to 
distinguish between two subclasses of clusters  shown in Fig. \ref{fig:clusters}. 
In type ``3a'' two of the sites can be simultaneously occupied, while in type
``3b'' only one of the sites can be occupied since all are mutually closer
than the particle diameter.  

\begin{figure}

\resizebox{8cm}{!}{\includegraphics{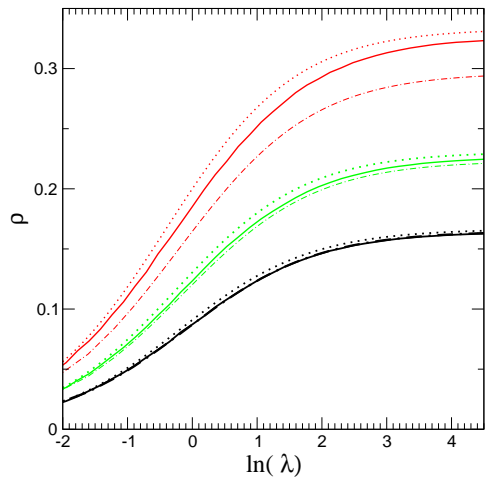}}
\caption{Adsorption isotherms versus activity $\lambda$ for several site densities: 
$\rho_s=0.5,0.3,0.2$ from top to bottom.   The dash-dotted and  dotted
lines show the    predictions of Eq.~\ref{eq:ce} (exact  expansion  to
third order in  $\rs$) and Eq.~\ref{eq:1}, respectively.   The solid lines
show the simulation results. }\label{fig:expan}.
\end{figure}

The predictions of Eq.~(\ref{eq:ce}) are compared with the simulation
results in  Fig.    \ref{fig:expan}.   The expansion to  third   order
provides a  good  description of the  isotherm  for $\rs=0.2$, but the
quality deteriorates rapidly for larger site density. Eq.~(\ref{eq:ce})
is unable to predict the correct limit when $\lambda \rightarrow
\infty$. This disagreement is more pronounced when $\rs$ increases:
for instance, when $\rs=1$, the expansion fails completely, because the
predicted density is lower than for $\rs=0.5$.

Since the above expansion  is limited to small  values of $\rho_s$, we
simplify Eq.~(\ref{eq:ce})   by    performing a    series expansion    in
$\rho_s$, giving

\begin{align}\label{eq:ce1}
\rho &= \lm \rho_s\bigl(\frac{1}{1+\lm}-\frac{2\lm}{(1+\lm)(1+2\lm)}\rs \nonumber\\
+& 
\frac{\lm^2(12\lm^2+29\lm+9)}{2(1+\lm)(1+2\lm)(1+3\lm)(1+3\lm+\lm^2)}\rs^2 +O(\rs^3)\bigr)
\end{align}

We note that Eq. (\ref{eq:ce1}) can be expressed as
\begin{align}
\rho(\rho_s,\lambda)&=\sum_{l=1}^3\left(-
\left(\frac{-\lambda}{1+\lambda}\right)^l +F_l(\rs,\lambda)\right)\rho_s^l
\end{align}
where $ F_l(\rs,\lambda)$ represents the  remaining terms in the exact
expansion  and,  consistent  with these  terms, has  the  property that
$F_l(\rs,\lambda)\to   0$        when    $\lambda\to   0      $    and
$\lambda\to\infty$. This property has been verified order by order for
a random lattice model\cite{TTV07}, and only for the three first orders
for this  model.  If we partially   resum the
series, neglecting the remaining terms, we obtain that
\begin{align}\label{eq:1}
\rho(\rho_s,\lambda)&=\frac{\lambda\rho_s}{1+\lambda(1+\rho_s)}
\end{align}
Although the site density expansion, Eq. (\ref{eq:ce1}), is valid only
for small $\rho_s$, the partial resummation leads to an expression for
$\rho$ that is exact in the limit  of very large activities $\lambda$,
whatever the   value of   $\rho_s$.    It thus  provides   a  sensible
approximation based on the site density expansion.

Isotherms    given  by   Eq.     (\ref{eq:1})  are   plotted in   Fig.
\ref{fig:expan} (dotted curves).  Although  for very low site  density
the third order  expansion gives a  better estimation,  the quality of
Eq. (\ref{eq:1}) is    significantly better for $\rho_s=0.5$, and   it
always gives the saturation density exactly.  For larger values of the
site density, $\rho_s>1$, the approximation overestimates the adsorbed
amount  for  intermediate  activities.  This  is   expected  since the
neglected terms are non-zero for intermediate bulk activity.  Attempts
to obtain     a   better   resummation   of the      series expansion,
Eq. (\ref{eq:ce1}), were fruitless.  Therefore, we follow the approach
that   we    used    in    our  study  of      the   model   of  dimer
adsorption\cite{TTV07}, by introducing an effective activity.

\subsection{Effective activity approach}

Despite  the exact mapping found in  the  infinite activity limit, no simple
reasoning   can  be  used to   obtain  exact  results at   finite bulk
activity. We therefore  investigated several approximate methods,  the
most successful of which is based on an effective activity.

\begin{figure}

\resizebox{7cm}{!}{\includegraphics{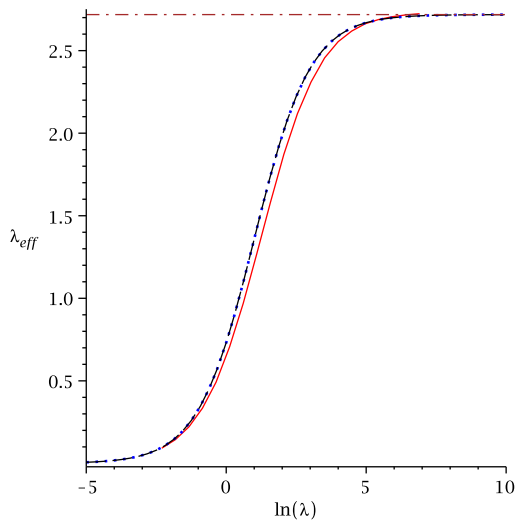}}
\caption{Effective activity $\lambda_{\rm eff}$ as a function of the logarithm of the activity $\ln \lambda$  with $\rs=1$. 
The   solid line shows  the   simulation result,  calculated from  the
density  by    using    Eq.  (\ref{eq:lam}).  The     predictions   of
Eq.~(\ref{eq:leffa})  combined    with         Eq.~(\ref{eq:3})    and
Eq.(\ref{eq:10}) are shown  as dotted and dashed lines,  respectively.
They are almost indistinguishable.   The horizontal dashed line  shows
the exact limiting value of $\rho_s e^{\rho_s}$.}\label{fig:effa1}.
\end{figure}

\begin{figure}
\resizebox{7cm}{!}{\includegraphics{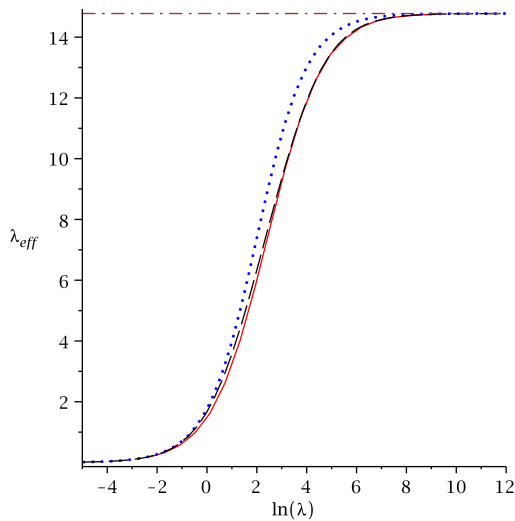}}
\caption{Same as Fig. \ref{fig:effa1}, except that $\rho_s=2$. The prediction of 
 Eq.~(\ref{eq:10}) is now  more accurate than that of Eq.~(\ref{eq:3}).}\label{fig:effa2}.
\end{figure}

\begin{figure}

\resizebox{7cm}{!}{\includegraphics{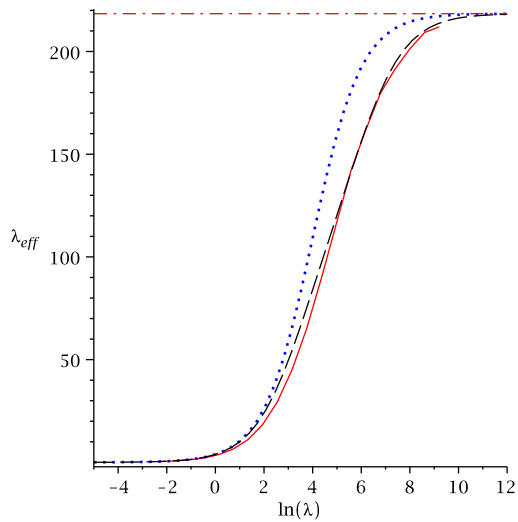}}
\caption{Same as Fig. \ref{fig:effa2}, except that $\rho_s=4$. 
Eq.~(\ref{eq:3}) fails  to reproduce  the simulation data  for intermediate
value of $\ln(\lambda)$ whereas the prediction
of Eq.~(\ref{eq:10}) remains accurate. }\label{fig:effa4}.
\end{figure}

The equation of state of the hard rod fluid on a continuous  line is \cite{T36}
\begin{equation}\label{eq:eos1d}
\beta P=\frac{\rho}{1-\rho\sigma}
\end{equation}
According to  the results  in  Section III,  the  hard rod  fluid at a
pressure $\beta P=\rho_s$ has the  same density (obtained by inverting
Eq. (\ref{eq:eos1d})) as the  densest configuration of hard spheres on
the random site surface, Eq. (\ref{eq:rmax}). We seek a generalization
of this mapping for an  arbitrary  value of  the bulk phase  activity,
i.e. an effective activity $\lm_{\rm  eff}$ such  that the density  of
adsorbed rods in the RSM, $\rho(\rho_s,\lambda)$ is given by
\begin{equation}\label{eq:leff}
\rho(\rs,\lm)=\rho^*(\lm_{\rm eff}(\lm,\rs)),
\end{equation}
where $\rho^*(\lm)$ is the density of rods on a continuous substrate at an
activity $\lm$. This is given exactly by
\begin{equation}\label{eq:rholam}
\rho^*(\lm)=\frac{L_w(\lm)}{1+L_w(\lm)},
\end{equation}
where $L_w(x)$, the Lambert-W function, is the solution of $x=L_w(x)\exp(L_w(x))$. 
The inverse relation is
\begin{equation}\label{eq:lam}
\lm = \frac{\rho}{1-\rho}\exp\bigl(\frac{\rho}{1-\rho}\bigr).
\end{equation}
From  the exact result,  Eq.  (\ref{eq:rmax}), we have that  $\lm_{\rm
eff}(\lm=\infty)=\rs e^{\rs}$.   Taking  Eqs.  (\ref{eq:rholam})  as a
mere definition of $\lm_{\rm  eff}(\lambda,\rho_s)$, we  have computed
it  from the simulated  adsorption  isotherms.  Results  are  shown in
Figs.~\ref{fig:effa1}-\ref{fig:effa4} and will be used as a reference
for testing the validity of various approximations.


Following  the method  introduced   in   the  study  of  the   lattice
model\cite{TTV07}, we attempt to construct simple expressions that can
reproduce   the numerical results,   when  available and  describe the
isotherms for a wide range  of parameters $\rho_s$  and $\lambda$.  The
effective activity as a function of the  activity $\lambda$ and of the
density site $\rho_s$ can be written as

\begin{equation}\label{eq:leffa}
\frac{1}{\lm_{\rm eff}}=\frac{f(\rho_s,\lm)}{\rs\lm}+\frac{1}{\rs e^{\rs}}
\end{equation}
where $f(\rho_s,\lm)$ is a function to be determined.  We know that at
small $\lm$, the  exact behavior is  given by $\lm_{\rm  eff}=\rs\lm$,
which  imposes  that $f(\rho_s,\lm)$  goes  to  $1$.  Conversely, when
$\lambda$ is large, the effective activity  $\lm_{\rm eff}$ behaves as
$\rho_s e^{\rho_s}+O(\lm)$,   which   leads  to the   constraint  that
$\lim_{\lm\rightarrow
\infty}f(\rho_s,\lm)=A(\rho_s)$ where $A(\rho_s)$ is an unknown function of $\rho_s$. 
If $A(\rho_s)$  remains close to  one, the simplest choice consists of
choosing
\begin{equation}\label{eq:3}
f(\rho_s,\lm)=1
\end{equation}
for  all values of $\lambda$.   While this simple  choice gives a fair
agreement   with the  simulated   values  when  $\rho_s=1$  (see  Fig.
\ref{fig:effa1}),   the situation deteriorates    for larger values of
$\rho_s$  (see    Figs.~\ref{fig:effa2}-~\ref{fig:effa4}).    However,
contrary to  the  one-dimensional lattice model,  we  do not know  the
exact asymptotic behavior of $f(\rho_s,\lm)$ for  large $\lambda$.  In
order  to improve the description,  we consider a homographic function
of  $\lm$ verifying the  exact behavior, when  $\lm\rightarrow 0$ and
$\lambda\rightarrow\infty$,
\begin{equation}\label{eq:4}
f(\rho_s,\lm)= \frac  {A(\rs) \lambda+B(\rs)}{ \lambda+B(\rs),
}
\end{equation}
where $A(\rs)$ and $B(\rs)$ are unknown functions.  Examination of the
series expansion in the limit of infinite  bulk activity suggests that
$B(\rs)$ should    grow as $e^{\rho_s}$ in   order  to give  the right
behavior.  In the  absence of additional  criteria,  we have chosen
 $A(\rs)=\rs$ and $B(\rs)=\rho_se^{\rs}$, which gives
\begin{equation}\label{eq:10}
f(\rho_s,\lm)= \frac {\rho_s(e^{\rho_s}+\lambda)}{  \lambda+\rho_s e^{\rho_s} }.
\end{equation}
Note that when $\rs=1$,  this function gives $f(\rho_s,\lm)=1$.  Figs.
\ref{fig:effa1}-~\ref{fig:effa4}  show   that  the   quality   of  the
approximation given  by  Eq.  (\ref{eq:leffa})  and  Eq.  (\ref{eq:3})
deteriorates with  increasing $\rho_s$.  This  is a consequence of the
fact  that the asymptotic   approach  to the saturated  state   is not
correctly captured with this simple approximation.  A better agreement
with   simulations is   obtained  by   using  Eq.(\ref{eq:leffa})  and
Eq.(\ref{eq:10})  for $\rs=2$  and $\rs=4$  (see Figs. \ref{fig:effa2}
and \ref{fig:effa4}).

\begin{figure}[t]

\resizebox{7cm}{!}{\includegraphics{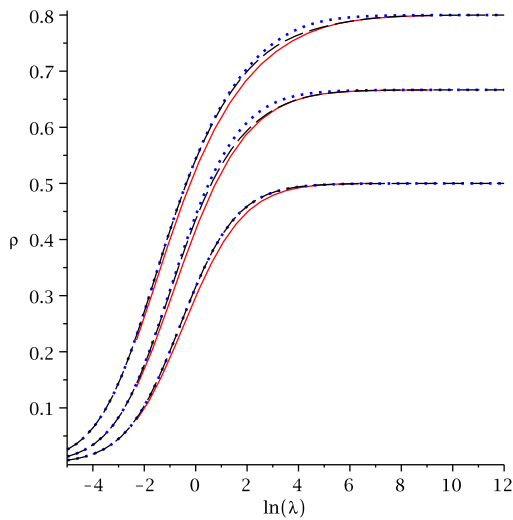}}
\caption{Adsorption isotherms versus activity $\lambda$ for several site densities: $\rho_s=1,2,4$, 
predicted by the effective activity  approach, Eqs. (\ref{eq:lam}) and
(\ref{eq:rholam})  by  using Eq.     (\ref{eq:3}) (dotted  lines)  and
combined with Eq.  (\ref{eq:10}) (dashed lines).  The solid lines show
the simulation results. }\label{fig:leff}.
\end{figure}

The  purpose of this exercise is  ultimately to obtain an approximate,
but  accurate, description of  the adsorption isotherms.  This is done
by  substituting    Eq. (\ref{eq:leffa})  in  Eq.   (\ref{eq:rholam}).
Fig.~\ref{fig:leff}  shows  a comparison of  these  estimates with the
simulation results.  The effective activity approach is a considerable
improvement over the cluster expansion,  whose accuracy is  restricted
to small density  site $\rho_s<0.5$ (c.f.   Fig.  \ref{fig:expan}) and
allows one to describe adsorption even when $\rho_s>1$.

\subsection{Structure}

\begin{figure}[th]

\resizebox{8cm}{!}{\includegraphics{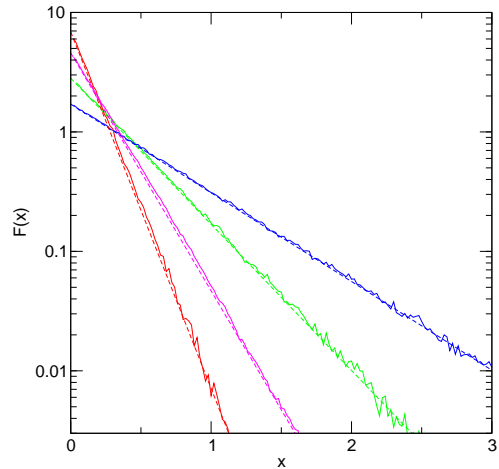}}
\caption{Gap distribution function for configurations of hard rods at finite activity
 on the random site surface with $\rs = 10$.  The dashed lines are the
 predictions  of Eq.   \ref{eq:gdist} with   a  density given  by  the
 equilibrium isotherm value and  the  solid lines show the  simulation
 results.   $\lm=1, 5, 50,  1000$  from  left to  right in the
 bottom part.}\label{fig:gapdist2}.
\end{figure}

We have also examined the  structure of the hard-rod configurations at
finite activity.  As in the case of the lattice model\cite{TTV07}, we
do not expect an  exact mapping to the homogeneous  system at the same
density.  Nevertheless, as  shown  in  Fig.  \ref{fig:gapdist2},   the
distributions computed from the simulation  are very well described by
Eq.  (\ref{eq:gdist}) with a density equal to the equilibrium isotherm
value.

\section{Two-dimensional model}

We  construct approximate isotherms  in the same way as 
for   one-dimensional model,  i.e.   by  introducing an effective
activity as a function of bulk   activity.  

\begin{figure}[t]
\resizebox{8cm}{!}{\includegraphics{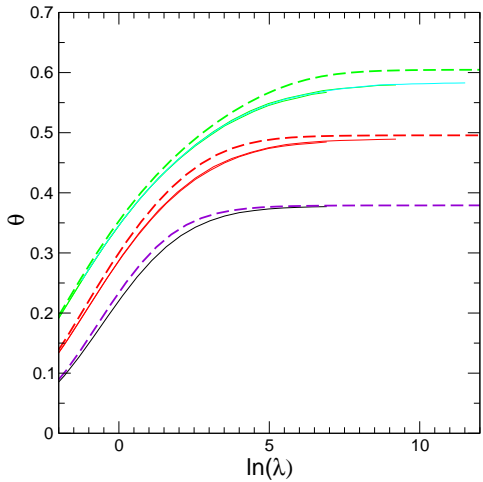}}
\caption{Adsorption Isotherms for dimensionless site densities $\alpha=1,2,4$ from bottom to top. The dashed  lines 
show the predictions of Eq.(\ref{eq:8}) combined with Eq.(\ref{eq:9}).
The solid lines show the simulation results obtained with 20 realizations of $n_s=2000$ sites. }\label{fig:rss2d}.
\end{figure}

We first consider  the homogeneous hard-disk system. We determine the
Helmholtz free energy per particle by integration along the isotherms:
\begin{equation}\label{eq:5}
\beta f^{ex}=\int_0^\theta \left(\frac{\beta P}{\rho}-1\right)\frac{d\theta'}{\theta'}.
\end{equation}
The excess chemical potential can then be obtained from
\begin{equation}\label{eq:6}
\beta \mu^{ex}=\beta f^{ex}+\theta\frac{\partial\beta f^{ex}}{\partial \theta}.
\end{equation}
By inserting the approximate equation of state, Eq. (\ref{eq:wang}) in Eq.(\ref{eq:5}), and substituting the result in Eq. (\ref{eq:6}), one obtains for the activity, $\lambda=\exp(\beta \mu)$ the relation
\begin{align}\label{eq:8}
\ln(\lambda)=\ln(4\theta/\pi) -{\frac {96}{5}}\,e{\theta}^{5}-
20\,d{\theta}^{4}-{\frac {32}{3}}\,c{\theta}^{3}-6\,b{\theta}^{2}-4\,a\theta\nonumber\\-  D
  \ln  \left( 1 -\frac{2\,\sqrt {3}}{\pi}\theta \right) +2\,{\frac {  D
  \sqrt {3}\theta}{\pi -2\,\sqrt {3}\theta}},
\end{align}
where, we recall, $\theta=\pi\sigma^2\rho/4$.

The  isotherm   for the $2D$  RSM   can be calculated   by following a
procedure similar to that used in  the one-dimensional case.  By means
of Eq. (\ref{eq:bp}), one obtains the saturation
coverage   when  the bulk activity  is   infinite, $\theta_\infty$. Then, by
inserting this  quantity in  Eq.~(\ref{eq:8}), the effective activity 
$\lambda_{\rm eff}^\infty\equiv \lambda_{\rm eff}(\theta_\infty)$ is obtained.

The   counterpart   of
Eq. (\ref{eq:leffa})  for the two-dimensional system is now
\begin{equation}\label{eq:9}
\frac{1}{\lm_{\rm eff}}=\frac{1}{\rho_s\lambda}+\frac{1}{\lm_{\rm eff}^\infty}
\end{equation}
where, in the absence of more information, we have set $f(\rho_s,\lambda)=1$.

Therefore,  at  a given  bulk    activity $\lambda$ and site density $\rho_s$, one  obtains  an
effective activity by   using  Eq. (\ref{eq:9}), and by  inserting   the
effective activity in  Eq. (\ref{eq:8}), one  deduces the corresponding
coverage in the 2D RSM.

Fig.~\ref{fig:rss2d}    compares  the  simulation  results   and the
approximate isotherms for different values of the site density.  While
the     agreement is not perfect,  the     scheme  represents a  major
improvement compared  to the pertubative approach (third-order density
expansion) for  describing   situations  where  the density   site  is
moderate to  high and allows one to  develop new  approximate isotherm
equations.
\section{Conclusion}

We have presented a theoretical  and numerical study of the reversible
adsorption  of  hard    spheres on  randomly    distributed adsorption
sites. Unlike the case of irreversible adsorption,  there is no simple
mapping between the RSM and a system of  hard spheres on a homogeneous
surface. We have, nonetheless, been  able to obtain some exact results
in various limits: small and  large bulk activity  and small and large
site density.  We have proposed  an effective activity approach, which
interpolates between  the   known behavior, to obtain   an approximate
description  for  intermediate situations.   The adsorption  isotherms
predicted by this approach are  in excellent agreement with simulation
results. In  the two-dimensional case, we  have not investigated phase
transitions in  detail, although there does  appear to be a solid-like
phase for sufficiently high  site density and  activity.  It will also
be interesting to  investigate  reversible adsorption on  a  substrate
when  a  second source   of   disorder  is  present:  distribution  of
adsorption site  energies. The present approach   should apply to this
case as well.

\appendix
\section{Random Site Model: 
beyond  the  continuum limit}\label{sec:random-site-model}   The sites
being    uncorrelated,   Eq.(\ref{eq:15}), one   easily   obtains the
connected    $n$-site     correlated    functions   $\left(\overline{\eta({\bf
r}_1)\eta({\bf r}_2)...\eta({\bf r}_n) }\right)_c$ which read
\begin{equation}
\left(\overline{\eta({\bf r}_1)\eta({\bf
r}_2)...\eta({\bf r}_n) }\right)_c
=\rho_s\prod_{i<j}\delta({\bf r}_i-{\bf r}_j).
\end{equation}

The  $n$-site correlation functions  which  appear in Eq.(\ref{eq:pf})
can  be obtained from the above connected  functions by using an 
expansion in the (mean) site density:
\begin{align}\label{eq:17}
&\overline{\eta({\bf r}_1)\eta({\bf
r}_2)...\eta({\bf r}_n) }=\rho_s^n
+\rho_s^{n-2}\sum_{i< j} \left(\overline{\eta({\bf r}_i)\eta({\bf
r}_j) }\right)_c 
 \nonumber\\
 &+\rho_s^{n-3}\sum_{i< j< k} \left(\overline{\eta({\bf r}_i)\eta({\bf
 r}_j)\eta({\bf r}_k) }\right)_c 
\nonumber\\
 &+\rho_s^{n-4}\sum_{i< j < k < l} 
\left(\overline{\eta({\bf r}_i)\eta({\bf r}_j)\eta({\bf r}_k)\eta({\bf r}_l)) }\right)_c
  \nonumber\\
 &+\rho_s^{n-4}\sum_{i< j} \left(\overline{\eta({\bf r}_i)\eta({\bf r}_j) }\right)_c 
               \sum_{i<k< l} \left(\overline{\eta({\bf r}_k)\eta({\bf r}_l) }\right)_c
\nonumber\\
&+... 
\end{align}
The expression of the higher-order terms rapidly becomes tedious.

On the  other hand,  the $n$-body connected  density functions   of a homogeneous hard sphere system at activity $z$ 
are  given  by functional  derivatives of the the logarithm of  grand-canonical
function (see Eq.(\ref{eq:11}))
\begin{align}\label{eq:16}
&\rho^{(p)*}_c({\bf                 r}_1{\bf                      r}_2...{\bf
r}_p|z)=\left.\frac{\delta^p\ln(\Xi^*(z({\bf    r})))}{\delta     z({\bf
r}_1)\delta  z({\bf  r}_2)...\delta z({\bf r}_p)}\right|_{z({\bf r})=z
}\nonumber\\      &=\sum_{n=p}^\infty\frac{z^n}{(n-p)!}\int      d{\bf
r}_{p+1}...d{\bf r}_{n} U_n({\bf r}_1,{\bf     r}_2...{\bf     r}_p,{\bf
r}_{p+1}...{\bf r}_n).
\end{align}
By inserting   Eq.(\ref{eq:17})   in  Eq.(\ref{eq:11}) and   by  using
Eq.(\ref{eq:16}),  the  disorder averaged  logarithm of  the partition
function of the RSM has the following expansion:
\begin{align}\label{eq:20}
&\overline{\ln(\Xi)}=\ln(\Xi^*(z))+\frac{1}{2\rho_s}\int d{\bf r}_1 \rho_c^{(2)*}({\bf r}_1,{\bf r}_1|z)\nonumber\\
&+\frac{1}{3!\rho_s^2}\int d{\bf r}_1 \rho_c^{(3)*}({\bf r}_1,{\bf r}_1,{\bf r}_1|z)\nonumber\\
&+\frac{1}{4!\rho_s^3}\int d{\bf r}_1 \rho_c^{(4)*}({\bf r}_1,{\bf r}_1,{\bf r}_1,{\bf r}_1|z)\nonumber\\
&+\frac{1}{4\rho_s^2}\int d{\bf r}_1d{\bf r}_2 \rho_c^{*(4)}({\bf r}_1,{\bf r}_1,{\bf r}_2,{\bf r}_2|z)\nonumber\\
&+...
\end{align}

For hard spheres, the non-overlapping property leads to a simple expression for the n-body connected density function
\begin{equation}\label{eq:18}
\rho_c^{(p)*}({\bf r}_1,{\bf r}_1,..,{\bf r}_1,{\bf r}_1|z)=(-1)^{p-1}(p-1)!\rho^*(z)^p.
\end{equation}
A straightforward but more tedious calculation allows one to obtain
\begin{equation}\label{eq:19}
\rho_c^{(4)*}({\bf r}_1,{\bf r}_1,{\bf r}_2,{\bf r}_2|z)=4\rho^*(z)^2\rho_c^{(2)*}({\bf r}_1,{\bf r}_2)-2\rho_c^{(2)*}({\bf r}_1,{\bf r}_2)^2
\end{equation}
By inserting Eqs.(\ref{eq:18}) and (\ref{eq:19}) in Eq.(\ref{eq:20}), one obtains that
\begin{align}
&\frac{\overline{\ln(\Xi)}}{A}=\frac{\ln(\Xi^*(z))}{A}+\left(-\frac{\rho^*(z)^2}{2\rho_s}+\frac{\rho^*(z)^3}{3\rho_s^2}-\frac{\rho^*(z)^4}{4\rho_s^3}\right)
\nonumber\\
+&\frac{\rho^*(z)^4}{2\rho_s^2}\int d{\bf r}(2h^*({r}|z)-(h^*({r}|z)))^2
\end{align} 
where  $h^*(r|z)$ is the  radial pair  correlation correlation of  the
hard-sphere model   at  the  activity $z$\cite{HM76}.   It    is worth
mentioning that the  second   term  of the rhs    of  Eq.(\ref{eq:20})
corresponds to  successive terms of   a series expansion that  can  be
resummed, and finally, one obtains that
\begin{align}\label{eq:21}
&\frac{\overline{\ln(\Xi)}}{A}=\frac{\ln(\Xi^*(z))}{A}+\rho_s\ln\left(1+\frac{\rho^*(z)}{\rho_s}\right)-\rho^*(z)
\nonumber\\
+&\frac{\rho^*(z)^4}{2\rho_s^2}\int d{\bf r}(2h^*({r}|z)-(h^*({r}|z))^2)+O(1/\rho_s^3).
\end{align}



\end{document}